# Incorporating uncertainty into the study of animal social networks[1]

David Lusseau[1*], Hal Whitehead[1], Shane Gero[1]
[1]Dalhousie University, Department of Biology, 1355 Oxford Street, Halifax, NS B3H 4J1, Canada. * d.lusseau@dal.ca

Network analysis is rapidly establishing itself as a powerful tool for studying the structure and dynamics of complex systems (Albert & Barabasi, 2002; Newman, 2003). It has proven useful in understanding social interactions among humans and non-humans and how global properties emerge from them (Girvan & Newman, 2002; Watts et al., 2002; Dodds et al., 2003; Lusseau, 2003; Lusseau & Newman, 2004; Croft et al., 2005; Flack et al., 2006; Lusseau, 2007b). It has also been helpful in describing and predicting the behavior of technological networks and some biological systems for which all interactions can be described as known absolute values. However, the application of network analysis to social systems involving non-human organisms has been slower, because it has been difficult to infer the statistical and biological significance of observed network statistics and structures (Croft et al., 2005; Lusseau et al., 2006).

Two key aspects have presented difficulties. Firstly, in contrast to some human studies, analysts estimate social relationships among individuals, they do not know them, and often they estimate those based on quite limited data. Researchers estimate relationships by observing interactions or associations between individuals, ranging from behavioral events (such as grooming) to co-occurrence. They can then build relationship measures using interaction rates or association indices (Whitehead & Dufault, 1999). Yet these observations do not represent all the interactions occurring between individuals, they are a sample. Studies in animal network analyses have never discussed sampling uncertainty even though its consequences can greatly affect the results of such analyses when sample size, i.e. the number of times individuals are observed, is small. For example if two individuals are together 50% of the time and so have a true association index (Cairns & Schwager, 1987) of 0.5, if they were identified together 10 times the 95% confidence interval for the estimated association index is about 0.3-0.7 (Whitehead, 2008).

A second problem is that most network analyses of non-humans have focused on binary networks, in which relationships are defined as being either present or absent. The matrix that represents the network contains only ones (when two individuals are defined as associated) and zeros (when they are not). Researchers have used binary transformations of continuous matrices of interaction rates or association indices to describe animal social networks. These transformations require certain arbitrary manipulations which can be justified to varying degrees (Lusseau, 2003; Croft et al., 2005). For example, one might decide that association indices smaller than an arbitrary value (say 0.5) should indicate the lack of a relationship (assigned a value of zero in the binary matrix) and those greater than 0.5 as a relationship (assigned a value of one in the binary matrix). Another example is to define pairs of which the association index is greater than expected if interactions occurred by chance as relationships (ones) and others not possessing relationships (zeros). Authors largely ignore these manipulations

---

when considering the conclusions derived from the results of these studies. In addition, most of these animal social systems are densely connected and discarding information about the strength of relationships might significantly distort the interpretation of the network topology. In many non-human communities, all individuals associate with all other individuals at some rate, so with complete sampling and association used to indicate relationships the binary network would link all individuals to all others. Different sampling rates, and different criteria for judging a dyad linked, can greatly change the perceived structure of a network (Croft et al., 2005). Binary simplification can lead to wrong interpretations about the social structure of the population. It can also lead to inappropriate divisions when defining community structure using these networks. Finally, it can also lead to wrong inferences about the position of individuals within the network.

We can also define networks with links between individuals representing the weight of associations between those individuals. These weighted networks can represent the matrix resulting from observations of interactions between or associations among individuals in the wild. Recent advances in weighted network analyses provide new tools to quantify the position of individuals in weighted networks and the community structure of those networks (Barrat et al., 2004; Newman, 2004a; Newman, 2006b). In our view, these tools are particularly appropriate for the analysis of non-human social networks. However, a shift towards weighted networks in animal behavior requires tools to deal with sampling issues. Here we introduce bootstrapping techniques to incorporate sampling uncertainty when estimating weighted network measures. We also introduce techniques that randomize networks subject to constraints, to assess how data structure influences the observed statistical properties of networks. We use two examples to illustrate the value of these new techniques. First, we will determine the variation in network centrality measures between individuals within a small sperm whale social unit (*Physeter macrocephalus*). We will then apply these methods to assess the uncertainty surrounding community structure in the bottlenose dolphin (*Tursiops* sp.) population residing in Doubtful Sound, New Zealand (Lusseau, 2003). Finally, using this bottlenose dolphin social network, we will test how transitivity in association departs from random. These analyses were implemented in Matlab using the Socprog package which is freely available at http://myweb.dal.ca/~hwhitehe/social.htm (Whitehead, 2009).

**Defining weighted networks**

Non-human societies, ranging from social insects to mammals, are commonly studied using dyadic association data, that is observations of interactions between pairs of individuals (Whitehead, 1997; Whitehead & Dufault, 1999; McComb et al., 2000; Watts, 2000; Shimooka, 2003; Sigurjonsdottir et al., 2003; Boogert et al., 2006; Greene & Gordon, 2007; McDonald, 2007). Association measures should indicate whether a pair of animals are in circumstances in which they may behaviorally interact (Whitehead & Dufault, 1999), and are often based upon common membership of transitive groups or other symmetric measures (e.g., within x body lengths), but asymmetric association measures are possible (e.g., nearest neighbors). We will limit our explanation to the former type of data because the analysis of asymmetric association data requires further manipulations of network statistics that are beyond the scope of this study.

Analysts record associations among animals in sampling periods, and then use these data to calculate association indices (Cairns & Schwager, 1987) which vary from 0 (never found associated) to 1 (always found associated). The resulting association matrix is the basis of many traditional analyses of non-human social structures (Pepper et al., 1999; Whitehead & Dufault, 1999), and it also defines a weighted network. In a display of this network, nodes represent individuals, and linking edges have line widths proportional to the association index between the two individuals (e.g. Fig. 1).

**Incorporating uncertainty in centrality measures**

The patterns of interactions within small social communities are difficult to quantify because of the issues associated with statistical inference based on a small number of data points (individuals in this case). It can therefore be difficult to understand whether different individuals play different structural role within these units (Lusseau, 2007a). Sperm whales (*Physeter macrocephalus*) live in matrilineal populations and female sperm whales spend most of their life within their natal unit (Whitehead, 2003). However, the structure of social relationships within these social units is not clear (Christal & Whitehead, 2001). Matrilineal social units in sperm whales function to provide care for calves at the surface while mothers make deep dives for food (Whitehead 2003). As such, a calf should be a central focus of the unit's underlying social relationships to maximize the likelihood it will survive.

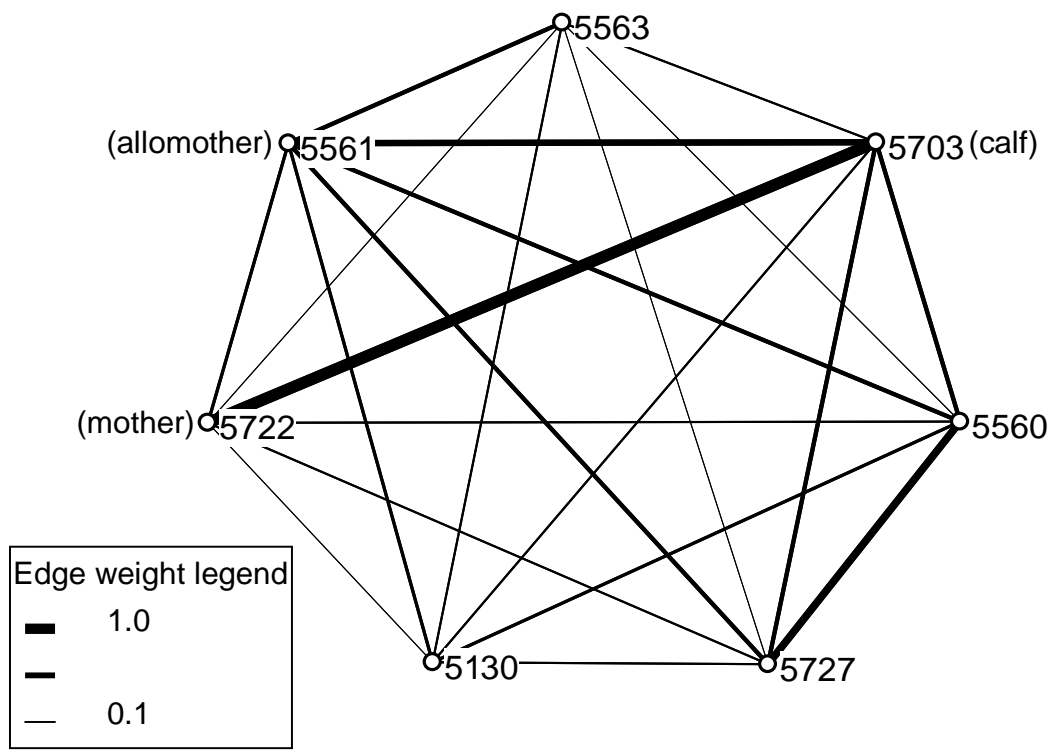

**Fig 1.** The association social network of the GOS social unit of sperm whales, the thickness of the lines (edges) represents the weight of the association index (half-weight index).

We examined this issue using data collected on a social unit, the *Group-of-Seven* (GOS), in an area that covered approximately 2000km$^2$ off the Commonwealth of Dominica (Gero, 2005). The GOS consists of five adult females, one juvenile male (#5727, 8-10 years old) and one male calf (#5703) whose mother was #5722. Following previous studies (Whitehead, 2003), we considered that individuals photo-identified together in clusters, defined as individuals within approximately 3 adult-body lengths from any other member and coordinated in their behavior. We used a half-weight association index (Cairns & Schwager, 1987) to define relationships. We collected 515 cluster samples over 72 days in January-March 2005 and 2006. From the network defined by the matrix of association indices (Fig. 1), we calculated centrality measures for each individual (Fig. 2 a-d). Several centrality statistics assess different aspects of the position of individuals within the network. Strength measures the general sociability of an individual and is the sum of its association indices (Newman, 2004a):

$$s_i = \sum_{i \neq j} AI_{ij}, \text{ where } AI_{ij} \text{ is the association index between i and j} \qquad [1]$$

However, this statistic does not provide any information about the range of associations. The coefficient of variation of association indices of an individual provides a measure of the heterogeneity of an individual's relationships (although this will be affected by sampling effort):

$$CV_i = \frac{sd(AI_{ij})}{\overline{AI_{ij}}} \qquad [2]$$

Eigenvector centrality is another measure of how well connected an individual is (Newman, 2004a). Mathematically, these centralities are simply elements of the first eigenvector of the matrix of edges or weights (e.g., an association matrix). They indicate the contribution of each individual to the structure of the association matrix. It indicates its connectedness within the network. Thus, an individual can have high eigenvector centrality either because it has high gregariousness or strength, or because it is connected to other individuals with high gregariousness. Finally, the clustering coefficient is helpful for understanding the transitivity of associations around an individual, i.e. the clustering coefficient for individual *a* indicates how well connected the individuals that are connected to *a* are to each other. Here we used a version for weighted networks introduced by Holme et al. (Holme et al., 2004):

$$c_i = \frac{\sum_j \sum_h AI_{ij} \cdot AI_{ih} \cdot AI_{jh}}{\max_{ij}(AI_{ij}) \cdot \sum_j \sum_h AI_{ij} \cdot AI_{ih}} \qquad [3]$$

It is not possible to test for the significance of the observed differences in centrality measures among individuals without an understanding of the confidence we have in those estimates. Bootstrapping can help us assess the confidence with which we estimate association indices and consequently derived network measures. In this re-sampling method the observation samples are considered to represent the best understanding we have of real associations between individuals in the population. We can obtain a bootstrap replicate of the data by re-sampling (with replacement) these samples. This replicate has the same sample size as the real data and we can obtain an estimation of the association matrix using these bootstrapped data. The process is then repeated a number of times, typically 1000, to obtain robust estimates of confidence intervals (Efron & Tibshirani, 1993). The confidence intervals surrounding each pairwise association index is then inferred from the observed variance in association indices in the bootstrap replicates (Efron & Tibshirani, 1993). Thus, we obtained 1000 bootstrapped association matrices for which we could measure network statistics and we used these to estimate confidence intervals for the statistics. While centrality statistics vary greatly between individual sperm whales within the social unit, their bootstrap errors largely overlap (Fig. 2 a-d). Pairwise comparisons show that some individuals do have significant differences; for example #5703, the calf, has a significantly greater contribution to the dominant eigenvector than any others (Fig. 2d). This analysis shows that some individuals have different contributions to the structure of the network. However, the clustering coefficient did not vary significantly among individuals of the sperm whale group (Fig. 2b).

This analysis is the first formal quantitative test showing that calves can play a significantly central role in the association patterns within the unit by being the individuals that contribute most significantly to the social network. Differences in centrality measures may also result from differences in the function of the association between individuals. Gero (2005) found that some individuals in this unit were more likely to escort (be in the same cluster at the surface) the calf than others. In particular, one individual, #5561, acted as the primary babysitter. A propensity for certain individuals to fill specific functional roles within the unit undoubtedly affects the structure of the social network.

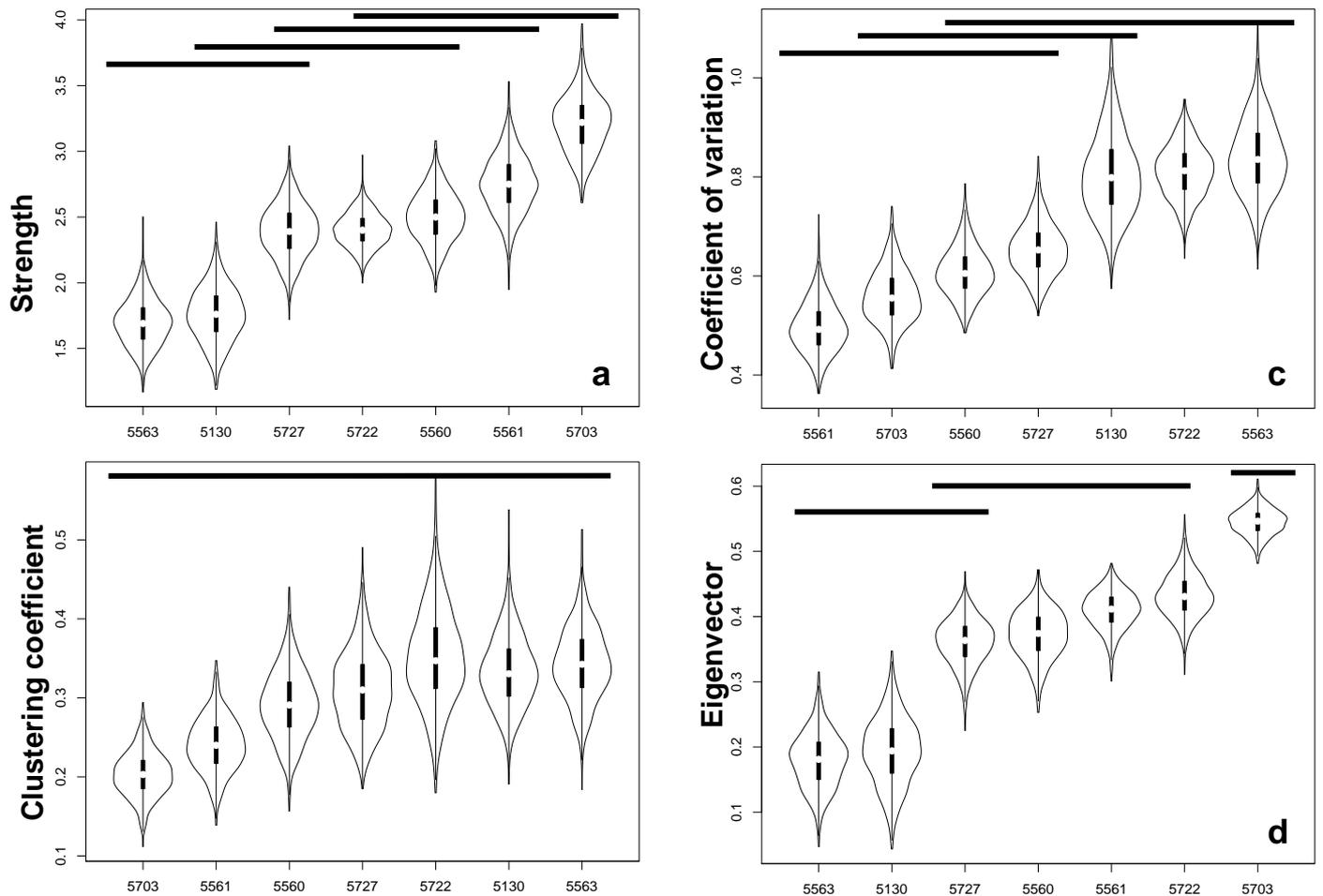

**Fig 2.** Violin plots of the bootstrapped (1000 iterations) centrality measures for each of the seven individuals: strength (a), clustering coefficient (b), Coefficient of variation of the association index (c), and contribution to the dominant eigenvector (d). Violin plots are composed of Kernel density estimates (Scott, 1992) (frequency distribution) mirrored on both sides of box plots for each individual. Plots were obtained using R (http://www.r-project.org). The 95% confidence intervals overlap was measured for pairwise comparisons.

## Defining community structure

One of the fundamental elements of the social organization of a group-living species is its community structure; that is how individuals segregate into communities in the population (Krause & Ruxton, 2002). This division is obvious in many instances because communities maintain clearly segregated home ranges. However, communities are not spatial segregated in many fission-fusion societies (Lehmann & Boesch, 2004; Croft et al., 2006; Lusseau et al., 2006; Ramos-Fernandez et al., 2006;

Sundaresan et al., 2007). Yet communities of individuals play a fundamental role in the ecology and sociality of these populations and we can attempt to define them using association patterns (Clutton-Brock et al., 1999; Lusseau et al., 2006). Techniques for dividing association matrices into clusters of closely associated individuals, borrowed mainly from multivariate statistics ("cluster analysis"), are plentiful but rarely provide consistent results (Whitehead & Dufault, 1999). There is also no well-accepted metric for comparing the acceptability of different clustering configurations.

We used a recently introduced network modularity technique to identify communities in social networks (Newman, 2006b; Newman, 2006a). This technique is based on defining a parsimonious division of the network which maximizes the number (and weights) of edges within communities and minimizes the number, and weight, of edges between communities. A good cluster division provides many edges within clusters and few between (Newman & Girvan, 2004). A modularity coefficient can quantify this. This coefficient, Q, is the sum of associations for all dyads belonging to the same cluster minus its expected value if dyads associated at random, given the strengths of the different individuals. This coefficient has the advantage of considering the possibility that all individuals belong to only one cluster. Therefore, the "best" clustering of a network is the division that maximizes Q.

Newman (Newman, 2006b) recently introduced a clustering algorithm which uses the modularity matrix: the weight (association index in our case) between two vertices minus the expected weight if weights were randomly distributed, which is related to the strengths of the two individuals involved in the pair compared with the overall sum of weights in the association matrix (Newman, 2006a). The eigenvector of the dominant eigenvalue of this matrix provides a good division into two clusters (positive versus negative values on this vector, see Newman (2006b) for more details). The technique is then used iteratively splitting clusters produced by the previous division, and the candidate community division is provided by the iteration that maximizes the modularity coefficient (Newman & Girvan, 2004; Newman, 2006b).

We applied this technique to school membership data obtained on a small resident population of bottlenose dolphins (*Tursiops* sp) which lives in Doubtful Sound, Fiordland, New Zealand (Lusseau et al., 2003).. We observed 437 schools over 126 days from December 1999 to April 2002 (Lusseau et al., 2003). We used a half-weight association index (Cairns & Schwager, 1987). Applying the modularity matrix technique to the association matrix provided the same division into two social units (Fig. 3) as defined previously using a binary social network of preferred companionships and a variety of clustering methods (Lusseau & Newman, 2004; Newman, 2004b; Newman, 2006a).

*Incorporating uncertainty in community structure*

We can examine the uncertainty surrounding the structure of a network using bootstrapping. We used the Newman modularity matrix technique on bootstrap replicates obtained using the technique described in the previous section. We could then determine how often individuals that are at the border of clusters were classified as belonging to one cluster or the other. The results of using this method on the Doubtful Sound bottlenose dolphin data highlights the great advantages of incorporating uncertainty in the estimation of community structure (Fig. 3, $Q_{max}$=0.1; 95% confidence interval: 0.088-0.12). It is important to understand social unit membership accurately in order to define the socioecology of this population based on observed social behaviors (Lusseau, 2007a). While there were two social cores, some individuals could belong to one unit or the other with varying degrees of probability. Indeed, some individuals were equally likely to belong to one or the other and therefore unit membership could not be resolved for them. The presence of the social cores reinforces the definition of social units in this dolphin population. The social structuring of dolphin populations has been difficult to assert because they do not always display the home range segregation observed in other species living in fission-fusion societies (Lusseau et al., 2006). This echoes similar challenges arising in other fission-fusion species (Ramos-Fernandez et al., 2006). This study confirms the relevance of social relationships in defining social communities within populations, which has important implications for their conservation (Lusseau et al., 2006), the study of cooperation (Lusseau, 2007b), and the evolution of behavior (Whitehead & Rendell, 2004).

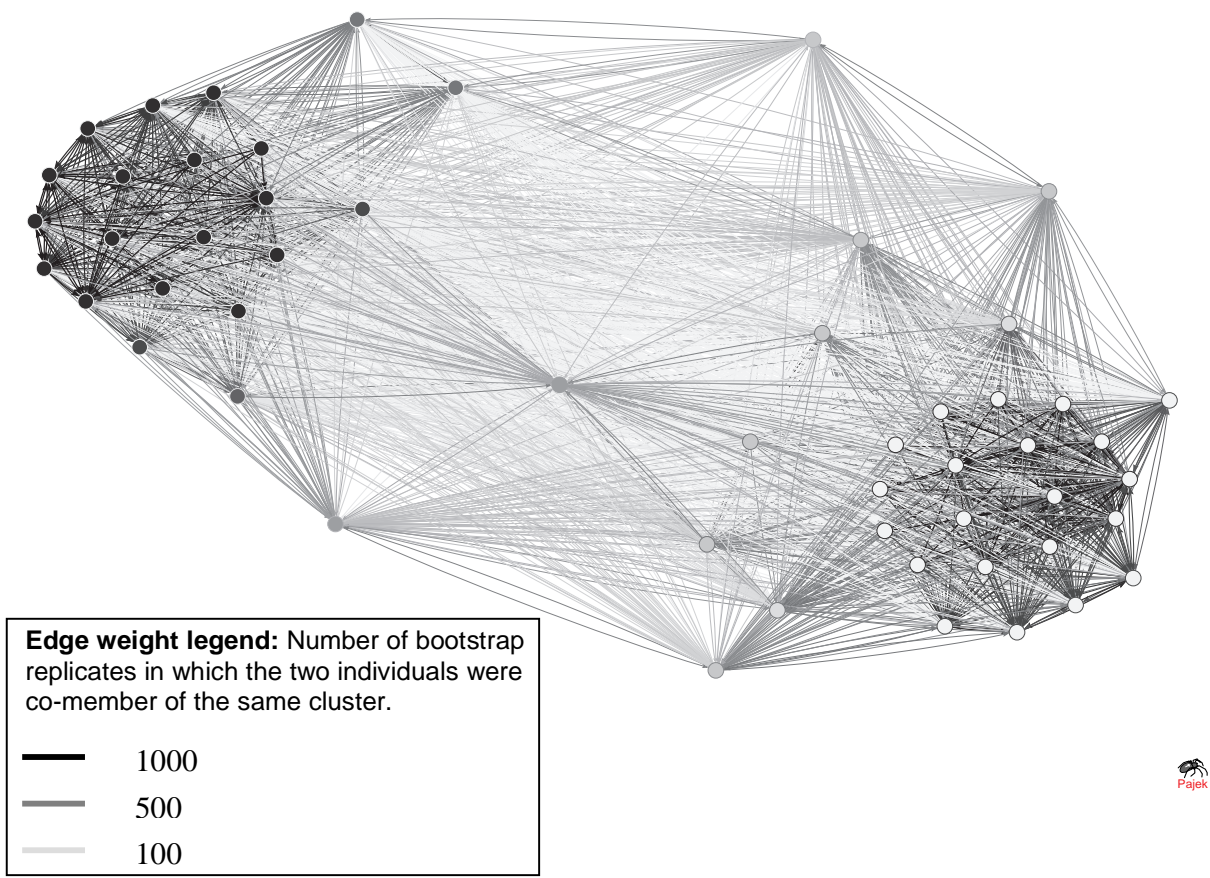

**Fig 3.** The Doubtful Sound bottlenose dolphin social network from associations observed from 1999 to 2002. Edges represent the number of times two individuals co-occurred in the same social unit over all bootstrap replicates. The shade of lines represents the likelihood of co-membership (the darker, the greater). The darker the center of a vertex, the more likely it is that the individual belongs to social unit 2. The darker the border of a vertex, the more likely it is that the individual belongs to social unit 1. The network was drawn using the Kamada-Kawai algorithm, a spring-embedder layout, (Kamada & Kawai, 1989) in Pajek (Batagelj & Mrvar, 2002).

**Understanding social behavior: randomization techniques**

Both social (i.e., attraction/avoidance of particular individuals) and gregarious (i.e., attraction/avoidance of individuals to other animals in general) behavior can contribute to observed association patterns; as can sampling. To understand the importance of social behavior in the observed association data, it is necessary to disentangle the contributions of social preferences, gregariousness, and sampling to the observed association indices. For example, clustering in animal society is an important measure for defining the openness of associations. Flack et al. (Flack et al., 2006) showed that in pigtailed macaques (*Macaca nemestrina*) the presence of policing individuals influenced the clustering coefficient of the society, promoting openness in interactions (a weaker average clustering coefficient than when policers were absent). Clustering coefficients can inform us about the likelihood that individuals associate with associates of their associates and therefore measure clustering. However, without having an understanding of what level of clustering we could expect given the gregariousness of a population, it is not possible to assess whether individuals do prefer to associate with the associates of their associates. While the Doubtful Sound bottlenose dolphin social network appear highly clustered, as there are many links (Fig. 3), much of this could relate to gregarious behavior since the small population lives in large schools (on average 17 dolphins in a school (Lusseau et al., 2003)).

    We can compare real data to that produced by making associations "random" to find out whether individuals do prefer to cluster in this population. However, randomizing networks is not as trivial as first thought (Amaral & Guimera, 2006). Erdös-Rényi random networks (in which links are laid down randomly) are often used but they may not always be appropriate because they do not account for the sampling structure of the data (Colizza et al., 2006). We used a modified version of the Bejder-Manly method, which is used to randomize association data, to obtain null random networks which control for the sampling structure and gregariousness of individuals (Manly, 1997; Bejder et al., 1998; Whitehead et al., 2005). The original method (Bejder et al. 1998) permutes group membership so that group size and the number of groups in which each individual was identified are both the same as in the original dataset. It does this by a series of flips in which randomly chosen records of individual A in group G and individual B in group H, are flipped to A in H and B in G (Manly, 1997).

    We compared the real weighted social network for the Doubtful Sound dolphin population to randomized counterpart networks produced using the Bejder-Manly technique to test whether social behavior, individual preferences, explained some of the observed clustering. We performed 1000 permutations with 100 flips per permutation (see (Whitehead et al., 2005) for more details), resulting in 1000 random networks. For the real network, and each of the random networks, we calculated the average clustering coefficient. We found that while the overall average clustering coefficient of the network did not differ from random ($c_{real}$ = 0.446, $c_{random}$ = 0.446, $p$ = 0.562), individuals from social unit 1 (white nodes in Fig. 3) did cluster a little more than expected by chance ($c_{real}$ = 0.450, $c_{random}$ = 0.448, $p$ = 0.013), while individuals from social unit 2 (black nodes in Fig. 3) clustered a little less than expected by chance ($c_{real}$

= 0.440, $c_{random}$ = 0.443, p = 0.001). Individuals belonging to different social units seem therefore to behave a little differently in the way they associate socially. Importantly, this analysis shows that aggregation explains most of the observed clustering. Since both social units occupy the same spatial range (Lusseau et al., 2003) and have similar age and sex class composition (Lusseau & Newman, 2004), this difference may only be explained by differences in behavioral preferences. Bottlenose dolphins can exhibit a range of diverse association behavior within and between populations (Connor et al., 1992; Lusseau et al., 2003). This analysis shows that clustering coefficient preferences can vary between social units within a population.

## Conclusions

These newly developed methods of analyzing weighted networks have considerable promise for the study of social networks, especially non-human societies, and in many ways complement traditional techniques. Weighted statistics provide a more realistic view of animal social networks. They also emphasize the diversity in relationships present in real data, which has proven extremely valuable in the study of human social networks (Onnela et al., 2007). Bootstrap and randomization techniques allow us to assess uncertainty relating to data structure and sampling in network statistics, at the level of the individual as well as the entire network. It is important to stress though that these techniques cannot substitute for insufficient data. They will assess the degree of confidence with which observed variation can be treated. Our results also emphasize the contributions of various factors to observed association rates that need to be considered when assessing the social relevance of network statistics.

## Acknowledgements

DL was supported by a Killam postdoctoral fellowship. Data collection and compilation for the Doubtful Sound study was funded by the New Zealand Whale and Dolphin Trust, the New Zealand Department of Conservation, Real Journeys Ltd, and the University of Otago (Departments of Zoology and Marine Sciences and Bridging Grant scheme). DL would like to thank Susan M. Lusseau, Oliver J. Boisseau, Liz Slooten, and Steve Dawson for their numerous contributions to the Doubtful Sound bottlenose dolphin data. Research in Dominica was carried out under scientific research permit (SCR 013/05-02) provided by the Ministry of Agriculture and Environment of the Commonwealth of Dominica and was supported by operating and equipment grants to HW from the Natural Sciences and Engineering Research Council of Canada and the Whale and Dolphin Conservation Society and to Luke Rendell from the U.K. Natural Environment Research Council. SG was supported by an NSERC Postgraduate Scholarship (PGSM). We would like to thank the Dalhousie University Department of Mathematics and Statistics for allowing us to use their computer cluster. We thank in particular Jeannette Janssen and Balagopal Pillai for their help with using the cluster.